# Production of light and intermediate mass fragments using various clusterization algorithms

*Ekta[1], Suneel Kumar[1] and Rajeev K. Puri[2]
*School of Physics and Materials Science, Thapar University Patiala-147004, Punjab India*
*Department of Physics, Panjab University, Chandigarh-160014, India*
* email: ektabansal76@gmail.com

## Introduction

In past years, a lot of efforts have been made experimentally and theoretically to understand the multifragmentation and its associated properties. It is well known that the colliding nuclei (at intermediate energies) shatter into several small, medium size pieces as well as lot of nucleons are also emitted which is known as multifragmentation. On the basis of theoretical scenario, one has the dynamical model where the reaction dynamics starts simulation from well defined nuclei to the end of the reaction where it is practically cold and scattered nuclear matter in the form of nucleons, light or heavy mass fragments. As a result no dynamical model simulates the fragments, rather one has the phase space of nucleons and constructs the fragments at the end of simulations. Therefore, we look for secondary models of clusterization algorithms e.g. minimum spanning tree (MST), minimum spanning tree with momentum cut (MSTP) and minimum spanning tree with binding energy cut MSTB [1].

## The Model

The present study is carried out within the framework of Isospin dependent Quantum Molecular Dynamical (IQMD) [2] which is based on event by event method & it treats the different charge state of nucleons, deltas and pions explicitly. The model is used to generate the phase-space of nucleons. The isospin degree of freedom enters through Coulomb potential, symmetry potential and NN cross section [2].

**Initialization:** The baryons are represented by Gaussian-shaped density distributions

$$f_i(\vec{r},\vec{p},t) = \frac{1}{\pi^2 \hbar^2} e^{-[\vec{r}-\vec{r}_i(t)]^2 \frac{1}{2L}} e^{-[\vec{p}-\vec{p}_i(t)]^2 \frac{2L}{\hbar^2}}$$

**Propagation:** The successfully initialized nuclei are then boosted towards each other using Hamilton equations of motion

$$\frac{dr_i}{dt} = \frac{d\langle H\rangle}{dp_i}; \quad \frac{dp_i}{dt} = -\frac{d\langle H\rangle}{dr_i}$$

With $\langle H\rangle = \langle T\rangle + \langle V\rangle$ is the total Hamiltonian.

$$\langle H\rangle = \sum_i \frac{p_i^2}{2m_i} + \sum_i \sum_{j>i} \int f_i(\vec{r},\vec{p},t) V^{ij}(\vec{r'},\vec{r}) \times f_j(\vec{r'},\vec{p'},t) d\vec{r} d\vec{r'} d\vec{p} d\vec{p'}$$

The total potential is the sum of the following specific elementary potentials.

$$V = V_{Sky} + V_{Yuk} + V_{Coul} + V_{mdi} + V_{loc}$$

**Collision:** During the propagation, two nucleons are supposed to suffer a binary collision if the distance between their centroid is

$$|r_i - r_j| \leq \sqrt{\frac{\sigma_{tot}}{\pi}}$$

Where $\sigma_{tot} = \sigma(\sqrt{s}, type)$

The collision is blocked with a possibility
$$P_{block} = 1-(1-P_i)(1-P_j)$$
Where $P_i$ and $P_j$ are the already occupied phase space fractions by other nucleons.

## Secondary Models (Different clusterization methods)

In minimum spanning tree (MST), two nucleons share the same fragment if their centroids are closer than a distance $d_{min}$ i.e. 4 fm

$$|r_i - r_j| \leq d_{min} \approx 4 fm \qquad (1)$$

where $r_i$ and $r_j$ are the spatial positions of both nucleons. Its influence on multifragmentation (at 200-300 fm/c) is observed to be small. By definition, this method cannot address the time scale of fragmentation [1]. In addition to equation (1) we also check the relative momenta of nucleons. Therefore, we say two nucleons must obey equation (2).

$$p_i - p_j \leq p_{Fermi} \quad (2)$$

where $p_{Fermi}$ is the average Fermi momentum of the nucleons bound in a nucleus at its ground state which is about 268 MeV/c. This definition, discards all those nucleons which are too far in their momentum space. This is known as minimum spanning tree method with a momentum cut (MSTP) method. The minimum spanning tree with binding energy check (MSTB), in this method pre-clusters obtained with MST method are subjected to binding energy equation.

$$\zeta_f = \frac{1}{N^f} \sum_{i=1}^{N^f} \left[ \sqrt{(\vec{p_i} - \vec{p_{cm}})^2 + m_i^2} - m_i + \frac{1}{2} \sum_{1 \neq j}^{N^f} V_{IJ} \right] < E_{BIND}$$

(3)

Here, we take $E_{bind}$ = -4.0 MeV if $N^f \geq 3$ and $E_{bind}=0$ [2]. In this equation $N^f$ represents the number of nucleons bound in a fragment and $p_{cm}$ is the centre of mass- momentum in a fragment.

## Results and disscussion

For the present analysis we simulate the reaction $^{129}Xe_{54}+^{197}Au_{79}$ at E=50 MeV/nucleon respectively[3]. This reaction is simulated at different impact parameters using hard equation of state. For each reaction 1000 events have been generated. The stored phase space is then analyzed by using MST, MSTP and MSTB algorithms. Fig.1, displays the time evolution of different fragments light mass fragments (2≤A≤4) (LMF's) and medium mass fragments (5≤A≤9) (MMF's) and Intermediate mass fragments (5≤A≤54) (IMF's). In case of free nucleons using MST indicates 197 nucleons at t=0 fm/c which increase to 326 nucleons at 30 fm/c. A cluster consisting of 326 nucleons at 30 fm/c signifies that nuclear matter is compressed and there is high density phase. As two nuclei (i.e. target and projectile) have large relative momenta and compound nucleus is not stable and it decays by emitting light and intermediate mass fragments. In other words, we have an artificial phenomenon in MST. MSTB and MSTP identifies the free nucleons as early as possible. In the two cases, a check in the form of binding energy and momentum cut helps to identify the fragments quite early. The normal MST takes quite a long time to identify the stable fragments which are residual of excited fragments.

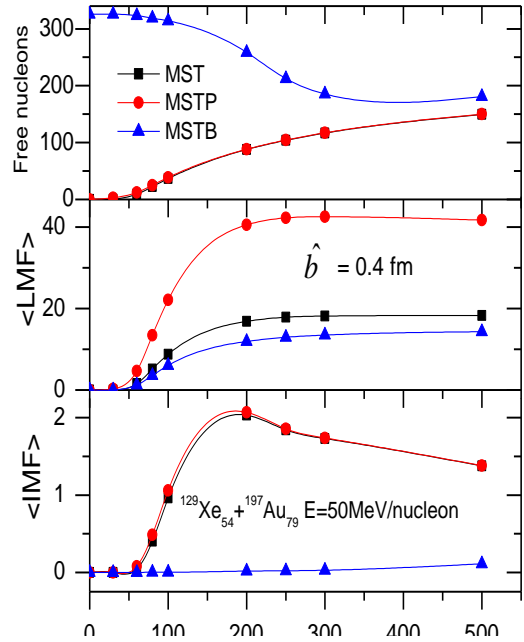

Fig1: The time evolution of different fragments with MST, MSTP and MSTB method.

## References:

[1] S. Kumar and R. K. Puri Phys. Rev. C **58**, 2858 (1998); ibid **58**, 320 (1998).
[2] C. Hatrtnack et al., Phys. J. **A 1** 151-169(1998).
[3] L. Phair et al., Phys. Lett. **B 285**, 10(1992).